# Investigation on spectral behavior of Solar Transients and their Interrelationship

Sharad C Tripathi[1*], Parvaiz A Khan[1], Aslam A M[1], A K Gwal[1], P K Purohit[2], Rajmal Jain[3]

1. Space Science Laboratory, Department of Physics, Barkatullah University, Bhopal – 462 026, M.P. (India)
2. National Institute of Technical Teachers' Training and Research, Bhopal – 462 002, M.P. (India)
3. Physical Research Laboratory, Navrangpura, Ahmedabad – 380009, Gujrat (India)

*Email: risingsharad@gmail.com

We probe the spectral hardening of solar flares emission in view of associated solar proton events (SEPs) at earth and coronal mass ejection (CME) acceleration as a consequence. In this investigation we undertake 60 SEPs of the Solar Cycle 23 alongwith associated Solar Flares and CMEs. We employ the X-ray emission in Solar flares observed by Reuven Ramaty Higly Energy Solar Spectroscopic Imager (RHESSI) in order to estimate flare plasma parameters. Further, we employ the observations from Geo-stationary Operational Environmental Satellites (GOES) and Large Angle and Spectrometric Coronagraph (LASCO), for SEPs and CMEs parameter estimation respectively. We report a good association of soft-hard-harder (SHH) spectral behavior of Flares with occurrence of Solar Proton Events for 16 Events (observed by RHESSI associated with protons). In addition, we have found a good correlation (R=0.71) in SEPs spectral hardening and CME velocity. We conclude that the Protons as well as CMEs gets accelerated at the Flare site and travel all the way in interplanetary space and then by re-acceleration in interplanetary space CMEs produce Geomagnetic Storms in geospace. This seems to be a statistically significant mechanism of the SEPs and initial CME acceleration in addition to the standard scenario of SEP acceleration at the shock front of CMEs.

## Introduction:

Solar transients; Solar Flares, Coronal Mass Ejections (CMEs), Solar Energetic Particles (SEPs) are the drivers of the Space Weather Effect in Geo-Space. There is a great discussion, in the community working in Space Weather research, on the association of these solar transients with each other. Sun in itself is a natural laboratory which provides us an opportunity to study the acceleration process of charged particles up to MeV- GeV energies. Solar Energetic Particles can escape into interplanetary space through open field lines and can be observed with in situ particle detectors, allowing the sampling of particles accelerated at Sun. The energy release through X-rays in solar flares is mostly due to bremsstrahlung emission. Good correlation has been found between spectral hardness of nonthermal HXR emission and X-ray flux at the corresponding energy (Grigis & Benz 2004; Fletcher & Hudson 2002) with initially soft spectrum before the peak flux, becoming harder as the flux increases, and becoming again soft as the flux decays, following a pattern soft-hard-soft (SHS) behavior. SHS behavior is thought to be a result of the electron acceleration mechanism in solar flares (Grigis & Benz 2006). Other possible causes are propagation effects of electrons traveling along flare loops and return currents caused by self-induced electric fields (Zarkova & Gordovskvy 2006). Instead of having SHS behavior some flares exhibit a HXR spectrum that successively hardens throughout flux peaks (Frost & Denis 1971; Cliver et al. 1986) and known as soft-hard-hard (SHH) behavior. SHH behavior of HXR emission is thought to be due to trapping of energetic electrons, with high-energy electrons being trapped the longest and found frequently in gradual flares. Kiplinger (1995) found the occurrence of flares with SHH spectral evolution to be closely associated with the observance of high-energy interplanetary proton events. This association between spectral hardening and SEP production suggests a connection between the HXR producing electrons in the flare loops and escaping energetic protons on open field lines. This is rather puzzling especially if the CME shock front is indeed the main proton accelerator. The spectral analysis of the solar flares and their relationship with the solar energetic particles has been extensively investigated by Saldanha et al., 2008 and Grayson et al., 2009, which reinforce the relationship between progressive hardening seen in the flare HXR emission and the SEP production first noted by Kiplinger (1995). CMEs are mostly thought to have their impact on the earth's environment. The relationship between solar flares and CMEs is found to be very complex and extensive efforts have been made to understand the cause and effect between them but so far nothing has been obtained satisfactorily. Often these two phenomena occur in conjunction but the relationship doesn't seem to be one to one. Statistical studies have indicated that higher intensity events are more likely accompanied by a CME (Harrison 1995). Kahler (1992) found that CMEs originate in the explosive phase of the associated flares which are long decay events. Green et al. (2002) have shown their analysis of an X1.2 class flare which doesn't has CME associated with. At this stage there are three ideas of flare-CME relationship; Dryer (1996) says that Flares produce CMEs, according to Hundhausen (1999) Flares are a by-product of CMEs whether Harisson (1995) and Zhang et al. (2001) claims that Flares and CMEs are part of the same magnetic eruption process. On temporal correspondence between CMEs and flares Harrison (1991) has concluded that CME onset typically precedes the associated X-ray flare onset by several minutes. Instead, Hundhausen (1999) considered this observational fact not to be responsible for flares to produce CMEs. Anzer & Pnueman (1982) suggested that the reconnection which leads the flare and also forms post-flare loops can propel overlying loops as CMEs. Zang et al. (2001) found in the investigation of four CMEs in comparison with time evolution on GOES X-ray flares that CMEs started accelerating impulsively until the peak of the soft X-ray flare which is consistent with the findings of MacQueen and Fisher (1983) that flare associated CMEs are in general faster than other CMEs.

In the present work we aim to verify whether the relationship between the solar flares and SEP events is consistent with Kiplinger study, using Reuven Ramaty High-Energy Solar Spectroscopic Imager (RHESSI) observations (Lin et al. 2002). The high spectral resolution of RHESSI (~1keV) allows us to clearly separate the thermal continuum from the nonthermal component of the spectrum and to follow the evolution of the nonthermal spectral index throughout the flare. Using RHESSI flare observations and SOHO/LASCO observations of

CMEs we are trying to find the relationship between spectral behavior of flares and associated CME dynamics.

## Data and Analysis

For the present work we have taken all the proton events of the solar cycle 23 and some of 24[th] as well under consideration. We have selected these events according to the definition of NOAA Space Environment Services Center of proton events as the events with peak flux ≥ 10 pfu by making a survey on the database www.solarmonitor.org which provides the status of the proton events observed by GOES satellites. The events selection of the events for association of the solar transients; Solar Flares, Coronal Mass Ejections and Solar Energetic Particles has been made using the criterion described in Fig.1.

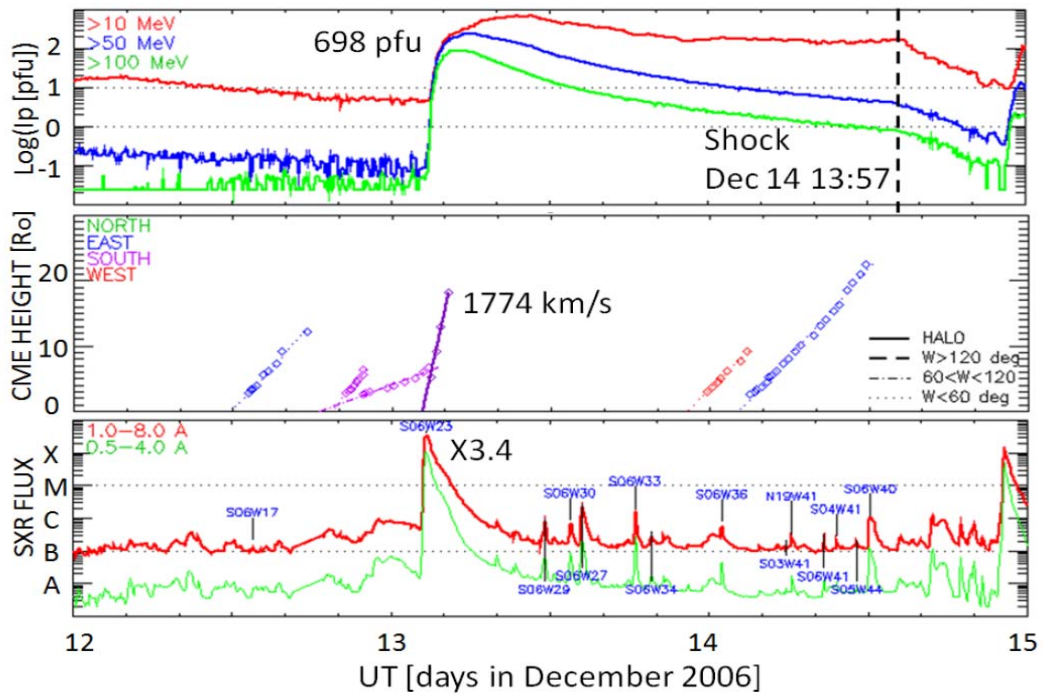

***Fig.1:*** *The gradual SEP event on 2006 Dec 13. Top panel shows Proton intensity in three energy channels (> 10 MeV, > 50 MeV, and > 100 MeV) with a maximum value of 698 particle flux units (pfu) in the > 10 MeV energy channel (1pfu = 1 particle per $cm^2$ .s.sr). The SEP intensity decreases in a rapid way after the shock arrived at 1 AU on December 14 at 13:57 UT (presented by the vertical dashed line). At the middle panel Height-time plots of all CMEs that occurred during 12-14 December 2006. The fast (1774 km/s) halo CME on December 13 at 02:54 was responsible for the SEP event and bottom panel shows GOES soft X-ray light curves in two energy channels (1.0-8.0 $A^0$ and 0.5-4.0 $A^0$). Several flares occurred during this period most of them coming from the active region NOAA 0930. The X3.4 flare associated with the SEP event occurred on December 13 at 02:40 UT.*

Then after we have taken data for all the events from the database SPIDR (http://spidr.ngdc.noaa.gov/spidr/) which provides proton data in different channels (0.8-4 MeV, 4-9 MeV, 9-15 MeV, 15-40 MeV, 40-80 MeV, 80-165 MeV, 165-500 MeV). Event integrated spectra of proton events has been made and fitted by power law, spectral indices of these spectrums are taken for the present analysis. Solar proton events occur when protons are accelerated to very high energies either close to the Sun during a solar flare or by Coronal Mass Ejection (CME) driven shock in the corona or in interplanetary space. We have taken the initial velocity of all the CMEs associated with the proton events from SOHO/LSCO catalogue at the website (http://cdaw.gsfc.nasa.gov/CME-list/). This catalog contains all the CMEs detected by the LASCO coronagraphs C2 and C3, which cover a combined field of view of 2.1 to 3.2 Rs. All the events selected for the analysis are tabulated in the table 1.

In order to study the evolution of solar proton events simultaneously with the dynamics of associated CMEs we have done some statistical analysis (after plotting the data points in order to enhance the S/N ration we binned the data in bins of 0.05 of proton spectral index) and found the linear relationship between the solar proton index and initial CME velocity with fairly good correlation coefficient (R= 0.71) by fitting the data points using linear fit (Fig.2). Correlation Coefficient 'R' is technically Pearson's correlation coefficient for linear regression. Binning has been done using different bin sizes of spectral indices and the comparative results are shown in the table 2.

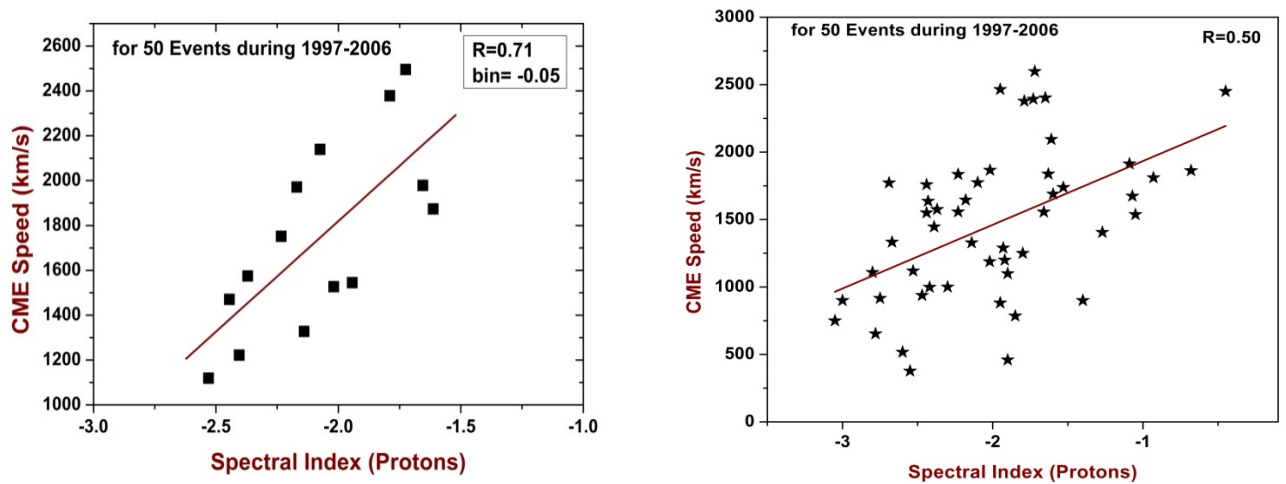

***Fig.2:*** *Relationship between Spectral Indices of 50 proton events during 1997-2006 and initial velocity of associated CMEs. (Left Panel: Spectral indices have been binned in bins of 0.05, Right Panel: All the events for a look of scatter).*

Following the criterion of event selection described above we have selected the solar flare events using RHESSI observations from 12 February 2002 (first data after launch) for the solar cycle 23-24. During the selection we have taken care that flares must have had at least partial observational coverage and we found 16 Flares observed by RHESSI associated with Proton Events during 2003-2006. RHESSI orbits Earth every ~96 minutes with ~36 minutes of night and occasional downtime due to the South Atlantic Anomaly due to which a significant fraction of solar activity is not observed by the satellite. In this way we have selected the events fully as well as partially observed by RHESSI. The RHESSI spacecraft is a light weight, NASA Small explorer mission and its low mass is partially due to lack of shielding around its detectors, resulting in relatively higher background radiation levels (Lin et al. 2002). RHESSI spectrograms (Krucker &Lin 2002) are individually inspected to ensure that there is enough nonthermal HXR emission above background levels to be analyzed by a power-law fit, since many of the smaller events exhibited mostly lower-energy thermal emission. We investigated each flare's spectral evolution to identify those with progressive non-thermal HXR hardening (SHH) behavior. For this purpose we used RHESSI front segment count rates, binned into 4 s time steps and covering ~ 3- 200 keV. We are interested in the relative behavior of the spectra through time so count rate data without full calibration have been used for the analysis. As the RHESSI response is almost linear in the relevant non-thermal HXR energy range of ~ 30-100 keV (Smith et al. 2002), it doesn't introduce significant bias. Before the fitting we have done the reduction of nighttime background subtraction. Pulse pileup has also been taken into account. In the events analyzed for the study hardness in the non-thermal energy range 50-100 keV is to be focused upon and non-thermal emission is considered to be significant above 50 keV energy range.  We have done the spectral analysis of 16 flares (written in bold in Table1) observed by RHESSI during the solar cycle 23 and associated with proton events in order to find temporal evolution of spectral index. The spectral analysis have been done using OSPEX package of Solar Soft for the appropriate subintervals and their respective spectra in the energy range 13-100 keV. For each flare interval non-thermal energy range was determined by S/N ratio to avoid the missing of the signal with the background. The spatially integrated count flux spectra have been fitted between 13-100 keV using the combination of isothermal component and a single power law. Most of the flares were best fitted with isothermal plus single power-law model as hard X-ray emission is produced by energized electrons via collisional bremsstrahlung,most prominently in the form of thick-target bremsstrahlung when precipitating electrons hit the chromospheres. Hard X-ray spectra can generally be fitted with a thermal spectrum at low energies and with a single or double power law nonthermal spectrum at higher energies. In some flares broken power law model has also been used if there is break (distinguished change in the slope) in the spectra at the higher order energies. For the present analysis temporal evolution of non-thermal spectral index have been explored in 50-100 keV energy range. The uncertainties in spectral index are those put by the OSPEX routines during the spectral fitting after setting the systemic uncertainty 0.00. The spectral evolution of the flare observed by RHESSI on 13 December 2006 is shown below in Fig.3 which shows SHH spectral behavior.

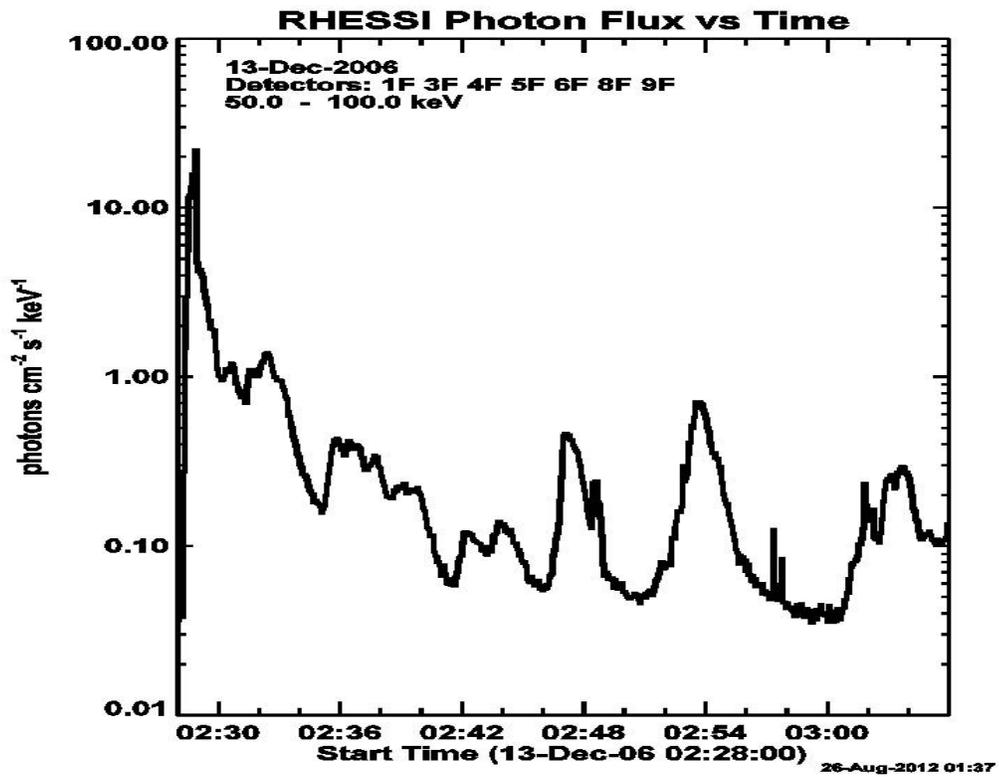

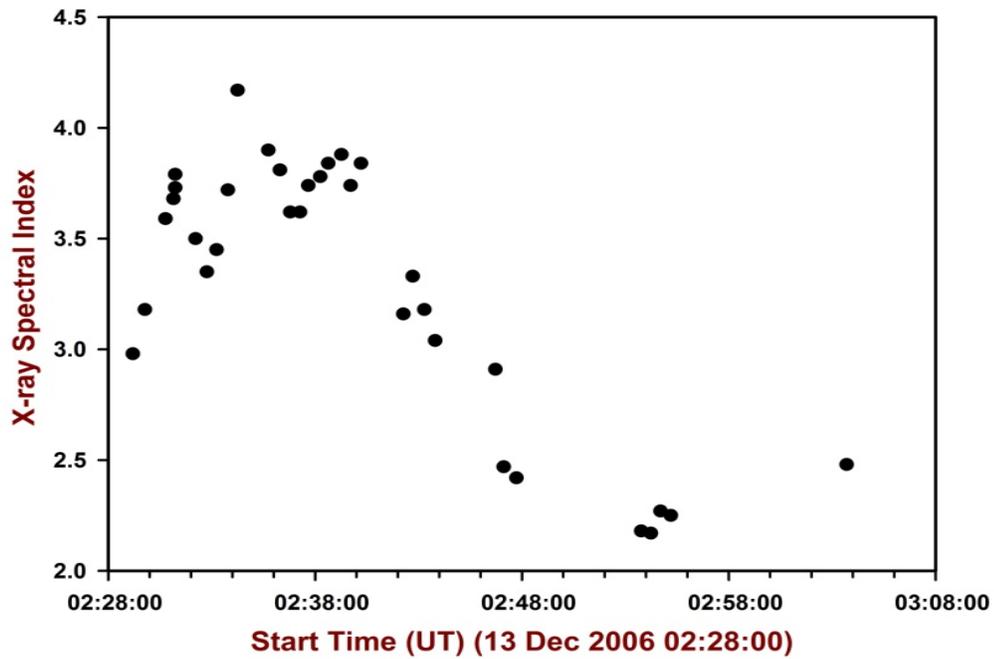

***Fig.3:*** *Top Panel: Temporal evolution of photon flux (50-100 keV) of flare observed by RHESSI on 13 December 2006 showing spectral hardening over time. Bottom Panel: Temporal evolution of the Spectral index of the the same flare.*

Spectral indices of the above flare events observed by RHESSI and that of associated proton events are found to have linear relation with each other with significantly fair correlation coefficient (R=0.8) Fig.4.

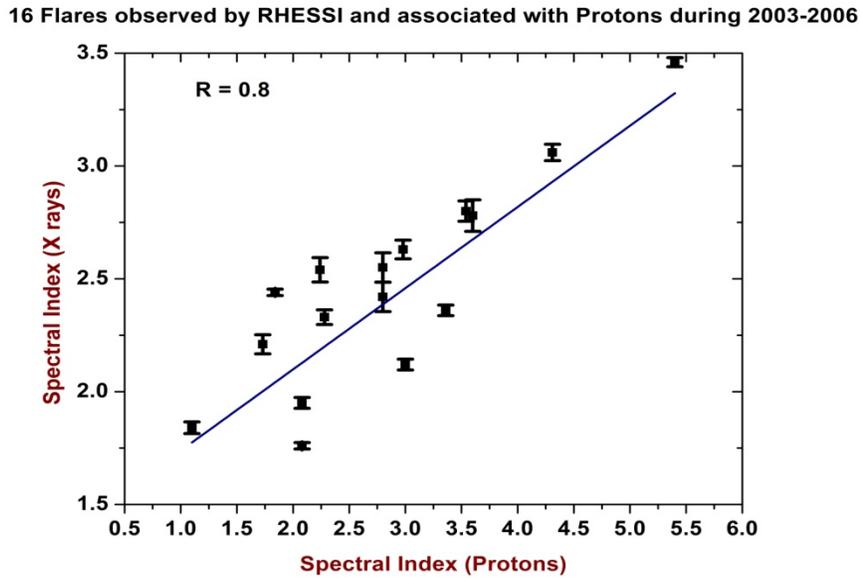

*Fig.4*: Relationship between Flare spectral index and associated Proton spectral index for 16 events during 1996-2006.

We have done rigorous spectral analysis of 32 events observed by RHESSI during 1996-2006 and found a power law relation between the spectral index (before the peak) of flares and initial speed of associated CMEs (R=0.53) Fig.5.

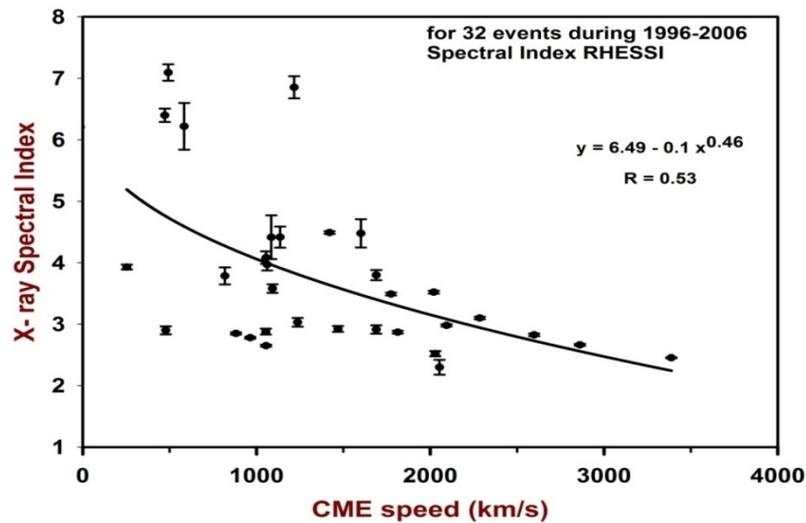

***Fig.5:*** *Relationship between Flare spectral index of 32 flares observed by RHESSI during 1996-2006 and initial velocity of associated CMEs (R=0.53).*

## Results and Discussion:

Following the analysis done in the present work we can infer about the relationship between the Solar Transients; Solar Flares, Coronal Mass Ejections, Solar Energetic Particles as these are seemed to be different manifestations of the same energy release processes on the Sun. Present study shows that as Hard X-ray emission is produced by energized electrons via collisional bremsstrahlung, most prominently in the form of thick-target bremsstrahlung when precipitating electrons hit the chromospheres, protons are also thought to be accelerated at the reconnection site and transported to all the way to earth through the open field lines. The linear relationship between proton and x-ray spectral indices shows that as the flare spectra harden proton spectra harden as well (Fig.4) which suggest that both the electrons producing X-rays and protons have the same origin in terms of genesis. Proton events are associated with the Solar Flares which have the SHH behavior as depicted in the Fig.3 for the flare observed by RHESSI on 13 December 2006 (Spectra hardens progressively with time). As mentioned in the table 1, 12 out of 16 RHESSI observed flares which are found to be associated with proton events show SHH spectral behavior. Fig.2 suggests that CME initial velocity is also dependent on the complex magnetic configuration at the corona and it seems to be accelerated at the reconnection site as well. CME velocity is also related with flare spectral index with the power law fit having correlation coefficient 0.53 (Fig.5) which shows the association of the flare and CME not only on temporal scale but also in terms of flaring plasma parameters (spectral index).

## Conclusion:

Solar transients; Solar Flares, CMEs and Solar Energetic Particles are the consequences of one energy release process in which the coronal magnetic energy releases in terms of flashes and mass motions. Solar Proton Events are mostly associated with the Solar Flares and thought to be accelerated at the reconnection site in the flaring plasma at the first sight as found in the correlative studies carried out in this paper (Fig.4) that there is a connection between these two solar transients, and being undergone through various acceleration processes through interplanetary medium as well. CME initial velocities are also found to be correlated with the X-ray as well as Proton spectral indices which shows that CMEs are being accelerated at the corona by one unique energy release process as do the flaring electrons and protons as well. The CME initial velocity seems to have some direct or indirect connection with the nature of the flaring plasma as all these solar transients: Flares, CMEs, Particles found to be produced from the magnetically complex regions in the solar corona. In this way the present study taking the spectral characteristics of the solar transients into consideration, infer positively that all these three types of solar transients do exist some relationship in their genesis.

## Acknowledgement:

Authors (Sharad C Tripathi and Parvaiz A Khan) are thankful to University Grant Commission for the financial support under Basic Scientific Research (BSR) Fellowship scheme during the work. Special thanks goes to Mr. K. J. Shah, Physical Research Laboratory for his help in OSPEX; RHESSI data analysis package of Solar Soft.**References:**

- Anzer, U., Pnueman, G. W. 1982, Solar Phys., 17, 129
- Cliver, E. W., Dennis, B. R., Kiplinger, A. L., Kane, S. R., Neidig, D. F., Sheeley, N. R., Jr., & Koomen, M. J. 1986, ApJ, 305, 920
- Dryer, M. 1996, Solar Phys., 169, 421
- Fletcher, L., & Hudson, H. S. 2002, Sol. Phys., 210, 307
- Frost, K. J., & Dennis, B. R. 1971, ApJ, 165, 655
- Grayson, J. A., Crucker, S., Lin, R. P. 2009, The Astrophys. J., 707, 1588-1594
- Green, L. M., Matthews, S. A., van Driel-Gesztelyi, L., Harra, L. K., Culhane, J. L. 2002, Solar Phys., 205, 2325–2339
- Grigis, P. C., & Benz, A. O. 2004, A&A, 426, 1093
- Grigis, P. C., & Benz, A. O. 2006, A&A, 458, 641
- Harrison, R. A. 1991, Adv. Space Res., 11(1), 25
- Harrison, R. A. 1995, A & A, 304, 585
- Hundhausen, A. J. 1999, Many Faces of the Sun (eds) Strong, K. T., Saba, J. L. R., Haisch, B. M., Springer-Verlag, New York, p. 143
- Kahler, S. W. 1992, Annual Review of A & A, 30 (A93-25826 09-90), 113–141
- Kiplinger, A. L. 1995, ApJ, 453, 973
- Krucker, S., & Lin, R. P. 2002, Sol. Phys., 210, 229
- Lin, R. P., et al. 2002, Sol. Phys., 210, 3
- MacQueen, R. M., Fisher, R. 1983, Solar Phys., 89, 89
- Saldanha, R., Krucker, S., Lin, R. P. 2008, The Astrophys. J., 673, 1169-1173
- Smith, D. M., et al. 2002, Sol. Phys., 210, 33
- Zhang, J., Dere, K. P., Howard, R. A., Kundu, M. R., White, S. M. 2001, Asrophys. J., 559, 452
- Zharkova, V. V., & Gordovskyy, M. 2006, ApJ, 651, 553

Appendix 1.

Table 1. Solar Transients during 1997-2006.

| S.N. | Date | Flare Onset Time GOES (UT) | GOES class | Flare Location | CME onset at C2 (UT) | CME linear speed (km/s) | >10 Mev proton onset (Date/UT) | Spectral Nature (RHESSI observation) |
|---|---|---|---|---|---|---|---|---|
| 1 | 1997Nov04 | 5:52:24 | X2.1 | S20W40 | 6:10:05 | 785 | 04/09:10 | |
| 2 | 1997Nov06 | 11:50:04 | X9.4 | S18W63 | 12:10:41 | 1556 | 06/12:00 | |
| 3 | 1998Apr20 | 9:09:15 | M1.4 | ? | 10:07:11 | 1863 | 20/12:00 | |
| 4 | 1998May02 | 13:28:30 | X1.1 | S17W24 | 14:06:12 | 938 | 02/13:55 | |
| 5 | 1998May06 | 7:57:00 | X2.7 | S11W65 | 8:29:13 | 1099 | 06/08:25 | |
| 6 | 1998Aug24 | 21:50:00 | X1.0 | N31E08 | data gap | data gap | 24/22:50 | |
| 7 | 1998Sep30 | 12:59:00 | M2.8 | N19W91 | data gap | data gap | 30/15:00 | |
| 8 | 1998Nov14 | 2:03:00 | C2.5 | ? | data gap | data gap | 14/06:00 | |
| 9 | 1999Jun01 | 18:53:00 | C1.2 | ? | 19:37:35 | 1772 | 01/21:00 | |
| 10 | 1999Jun04 | 6:52:00 | M3.9 | N20W85 | 7:26:54 | 2230 | 04/09:00 | |
| 11 | 2000Apr04 | 15:12:00 | C9.7 | N18W72 | 16:32:37 | 1188 | 04/15:30 | |
| 12 | 2000Jun06 | 14:58:00 | X2.3 | N21E10 | 15:54:05 | 1119 | 07/00:00 | |
| 13 | 2000Jun10 | 16:40:00 | M5.2 | N22W42 | 17:08:05 | 1108 | 10/18:00 | |
| 14 | 2000Jul14 | 10:03:00 | X5.7 | N17W11 | 10:54:07 | 1674 | 14/10:30 | |
| 15 | 2000Sep12 | 11:31:00 | M1.0 | ? | 11:54:05 | 1550 | 12/12:30 | |
| 16 | 2000Nov08 | 22:42:00 | M7.4 | N02W78 | 23:06:05 | 1738 | 08/23:55 | |
| 17 | 2000Nov24 | 4:55:00 | X2.0 | N21W07 | 5:30:05 | 1289 | 24/06:00 | |
| 18 | 2001Jan28 | 15:40:00 | M1.5 | ? | 15:54:05 | 916 | 28/16:00 | |
| 19 | 2001Apr02 | 21:32:00 | X20 | N16W70 | 22:06:07 | 2505 | 02/23:55 | |
| 20 | 2001Apr09 | 1:59:00 | C6.4 | S21W08 | 0:06:05 | 653 | 09/15:00 | |
| 21 | 2001Apr15 | 13:19:00 | X14 | S22W85 | 14:06:31 | 1199 | 15/14:00 | |
| 22 | 2001Apr18 | 2:11:00 | C2.2 | S23W117 | 2:30:05 | 2465 | 18/02:35 | |
| 23 | 2001Aug15 | 17:21:00 | C1.1 | ? | 23:54:05 | 1575 | 16/00:00 | |
| 24 | 2001Sep24 | 9:32:00 | X2.6 | S18E18 | 10:30:59 | 2402 | 24/12:00 | |
| 25 | 2001Oct01 | 4:41:00 | M9.1 | ? | 5:30:05 | 1405 | 01/12:30 | |
| 26 | 2001Nov04 | 16:03:00 | X1.0 | N05W29 | 16:35:06 | 1810 | 04/17:00 | |
| 27 | 2001Nov22 | 20:18:00 | M3.8 | S24W68 | 20:30:33 | 1443 | 22/21:00 | |
| 28 | 2001Dec26 | 4:32:00 | M7.14 | N08W54 | 5:30:05 | 1446 | 26/05:30 | |
| 29 | 2002Jan10 | 09/17:42 | M9.5 | N13W07 | 0:30:05 | 377 | 10/00:00 | |
| 30 | 2002Apr21 | 0:43:00 | X1.5 | S14W91 | 1:27:20 | 2393 | 21/02:00 | |
| 31 | 2002May22 | 3:18:00 | C5.0 | ? | 3:50:05 | 1557 | 22/06:00 | |
| 32 | 2002Jul16 | 13:31:00 | C8.5 | N18W14 | 16:02:58 | 1636 | 16/15:00 | |
| 33 | 2002Aug22 | 1:47:00 | M5.4 | S08W72 | 2:06:06 | 998 | 22/03:00 | |
| 34 | 2002Aug24 | 0:49:00 | X3.1 | S08W91 | 1:27:29 | 1913 | 24/01:18 | |
| 35 | 2002Sep05 | 20:40:00 | C8.6 | S16W01 | 21:56:08 | 517 | 06/03:00 | |
| 36 | 2002Nov09 | 13:08:00 | M4.6 | S10W42 | 13:31:45 | 1838 | 09/15:00 | |
| **37** | **2003May29** | **0:51:00** | **X1.2** | **?** | **0:50:05** | **1366** | **28/03:30** | **SHH** |
| **39** | **2003May31** | **2:13:00** | **M9.3** | **S07W73** | **2:30:19** | **1835** | **31/04:00** | **SHH** |
| **40** | **2003Jun17** | **22:27:00** | **M6.8** | **S07E57** | **23:18:14** | **1813** | **18/09:00** | **SHH** |
| 41 | 2003Oct26 | 17:21:00 | X1.2 | N04W41 | 17:54:05 | 1537 | 26/18:00 | |
| **42** | **2003oct28** | **11:00:00** | **X17** | **S16E04** | **11:30:05** | **2459** | **28/11:22** | **SHH** |
| **43** | **2003Oct29** | **20:37** | **X10** | **S17W10** | **20:54:05** | **2029** | **29/21:00** | **SHH** |
| **44** | **2003Nov02** | **17:03:00** | **X8.3** | **S17W63** | **17:30:05** | **2598** | **02/17:30** | **SHS** |
| 45 | 2004Jul25 | 13:37:00 | M2.2 | N08W35 | 14:54:05 | 1333 | 25/17:50 | |
| 46 | 2004Sep12 | 0:04:00 | M4.8 | N05E33 | 0:36:06 | 1328 | 13/20:30 | |
| 47 | 2004Sep19 | 16:46:00 | M1.9 | N06W59 | data gap | data gap | 19/18:00 | |
| **48** | **2004Nov01** | **3:04:00** | **M1.1** | **N13W42** | **3:54:05** | **459** | **01/06:00** | **SHS** |
| **49** | **2004Nov07** | **15:42:00** | **X2.0** | **N09W08** | **16:54:05** | **1759** | **07/16:00** | **SHH** |

| 50 | 2004Nov10 | 01:59:00 | X2.5 | N08W50 | 02:26:05 | 3387 | 10/03:00 | SHS |
| 51 | 2005Jan15 | 5:54:00 | M8.6 | N13W04 | 6:30:05 | 2049 | 15/07:00 | SHH |
| 52 | 2005Jan17 | 6:59:00 | X3.8 | N13W29 | 9:30:05 | 2094 | 17/09:00 | SHH |
| 53 | 2005Jan20 | 6:39:00 | X7.1 | N14W70 | 6:54:05 | 882 | 20/06:00 | SHH |
| 54 | 2005May13 | 16:05:00 | M8.0 | N12E05 | 17:12:05 | 1689 | 13/21:00 | SHH |
| 55 | 2005Jun16 | 20:01:00 | M4.0 | N09W88 | data gap | data gap | 16/21:00 | |
| 56 | 2005Jul27 | 04:33:00 | M3.7 | ? | 04:54:05 | 533 | 27/06:00 | SHS |
| 57 | 2005Aug22 | 16:46:00 | M5.6 | S11W62 | 17:30:05 | 2378 | 22/18:00 | |
| 58 | 2005Sep13 | 19:19:00 | X1.5 | S11E17 | 20:00:05 | 1866 | 13/23:00 | |
| 59 | 2006Dec05 | 10:18:00 | X9.0 | ? | data gap | data gap | 05/18:00 | SHH |
| 60 | 2006Dec13 | 2:14:00 | X3.4 | S06W35 | 2:54:04 | 1774 | 13/03:00 | SHH |

Table 2. Correlation coefficients between CME velocity and Proton spectral index for different spectral bin sizes.

| Bin Size (spectral index) | Correlation Coefficient |
|---|---|
| 0.03 | 0.54 |
| 0.04 | 0.64 |
| 0.05 | 0.71 |